\documentclass[prl,twocolumn,showpacs,amsmath,amssymb]{revtex4}

\usepackage{graphicx}
\usepackage{dcolumn}
\usepackage{bm}

\begin{document}

\preprint{APS/123-QED}

\title{Pattern Formation During Deformation of a Confined Viscoelastic Layer: \\From a Viscous Liquid to a Soft Elastic Solid.}

\author{Julia Nase}
 \author{Anke Lindner}
 \email{anke.lindner@espci.fr}
\affiliation{Laboratoire de Physique et M\'ecanique des Milieux H\'et\'erog\`enes (PMMH),UMR 7636
CNRS - ESPCI - Universit\'es Paris
6 et 7, 10, rue Vauquelin, 75231 Paris Cedex 05, France}%

\author{Costantino Creton}
\affiliation{Laboratoire de Physico-Chimie des Polym\`eres et des
Milieux Disperses (PPMD), UMR 7615 CNRS - ESPCI -Universit\'es Paris
6 et 7, 10 rue Vauquelin, 75231 Paris Cedex 05, France}%

\date{\today}

\begin{abstract}
We study pattern formation during tensile deformation of confined viscoelastic layers. The use of a
model system (PDMS with different degrees of crosslinking) allows us to go continuously from a
viscous liquid to an elastic solid. We observe two distinct regimes of fingering instabilities: a
regime called ''elastic`` with interfacial crack propagation where the fingering wavelength only
scales with the film thickness, and a bulk regime called ''viscoelastic`` where the fingering
instability shows a \textit{Saffman-Taylor-}like behavior. We find good quantitative agreement with
theory in both cases and present a reduced parameter describing the transition between the two
regimes and allowing to predict the observed patterns over the whole range of viscoelastic
properties.

\end{abstract}

\maketitle

\textsc{Introduction} -- Good soft adhesives show viscous and elastic properties that allow on the
one hand having a good molecular contact with the substrate and on the other hand a  resistance to
a certain stress level during debonding. The viscoelastic properties determine the debonding
mechanisms when being detached from a rigid substrate, involving the formation of complex patterns
as bulk fingering or interfacial crack propagation \cite{Shull_JPolScie_2004}. Pattern formation
during tensile deformation of thin layers in confined geometries has also attracted much interest
from a fundamental point of view. In the case of a purely viscous liquid confined between two
plates being separated, air penetrating from the edges leads to the formation of bulk fingers. This
fingering instability is well described by the classical \textit{Saffman -- Taylor} instability
\cite{Saffman_RROCRSOC_1958, Lindner_PHYSFLUIDS_2005, Derks_JAPPLPHYS_2003,
BenAmar_Physica_2005,Shelley_Nonlin_1997,Poivet_EPJE_2004}, where a less viscous liquid pushes a
more viscous liquid in a confined geometry. For a thin layer of a purely elastic material,
undulations of an interfacial crack front have been observed experimentally and explained
theoretically \cite{Ghatak_all,Adda-Bedia_PROCRSOCA_2006,Monch_Europhys_2001,Fields_PhilMag_1976}.
Some studies have focused on complex or yield stress fluids \cite{Derks_JAPPLPHYS_2003,
BenAmar_Physica_2005}, elastic gels \cite{Shull_PRL_2000,Webber_PHYSREVE_2003}, ferromagnetic
fluids \cite{Oliveira_PRE_2006}, pastes \cite{Lemaire_Fractals_1993}, or considered the role of the
substrate \cite{Sinha_EPJE_2008}. The transition between a viscous liquid and a glassy material has
been studied \cite{Harvey_JourOfMatSc_2003,Israelachvili_06_07}.

However no systematic study of the pattern formation during deformation of a viscoelastic material
focusing on the respective role of the liquid and elastic properties has been undertaken so far. We
present here a system involving a specifically designed model soft material with tunable properties
going continuously from a viscous liquid to an elastic solid. Studying the debonding mechanisms
using a probe tack test on these materials allows for the first time to explain the observed
patterns quantitatively over the whole range of viscoelastic properties and to describe the
transition between the two well known limits observed for a pure liquid or an elastic solid. Such a
study helps for a better understanding of the instabilities observed in the viscoelastic regime of
industrial applications. It is also of importance for any theoretical treatment aiming to bridge
the gap between the different formalisms that apply to viscous liquids and elastic solids.

\textsc{Materials and Methods} -- As model system we use a weakly cross linked polymer,
Poly(dimethylsiloxane) (PDMS). We chose the commercial product  \textit{''Sylgard\copyright\ 184
Silicone Elastomer Kit``} purchased at \textit{Dow Corning}. It consists of a silicone oil and a
curing agent that is able to form cross links, i.e. chemical bonds between the polymer chains. The
non cured silicone oil is a Newtonian liquid. Adding curing agent increases the number density of
cross link points and the material becomes viscoelastic. The fully cured PDMS at $10\%$ of curing
agent is an elastic solid. This system thus represents an ideal model system providing a
reproducible and easy way to go continuously from a viscous liquid to an elastic solid.

To determine the material's linear rheological properties, we perform oscillatory frequency sweep
tests after curing in a plate-plate geometry. This gives access to the storage and loss moduli $G'$
and $G''$ that are measures for the material's elastic and viscous properties, respectively, as
well as to the complex modulus $G^\star = \sqrt{{{G}^{'}}^2 + {{G}^{''}}^2}$.

\begin{figure}
    \centering
    \includegraphics[width=7cm]{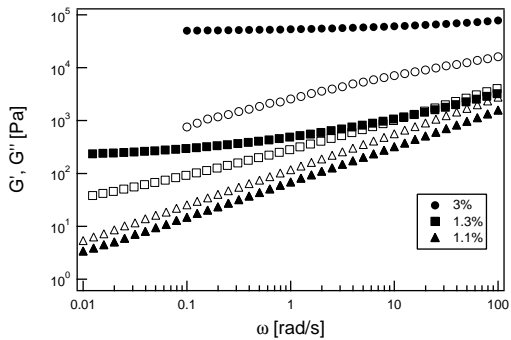}
    \caption{The storage and loss moduli $G'$ (full symbols)
    and $G''$ (open symbols) as a function of the frequency $\omega$ for different amounts of
    curing agent.}
    \label{fig:G_vs_freq}
\end{figure}

Figure \ref{fig:G_vs_freq} shows the results for different amounts of cross linker. The material
with about $3\%$ of cross linker is elastic, having a $G'$ several orders of magnitude higher than
$G''$; adding about $1\%$ of cross linker leads to a product in the viscoelastic regime close to
the gel point.

We prepare polymeric films on microscope glass slides ($10\times2.6\times0.2cm$) that are
precleaned and coated with a primer (Dow Corning 1200 OS) to enhance the adherence of PDMS to the
slide. We use applicators to deposit films of different thicknesses. The samples are cured in a
desiccator at $80^{\circ}C$ for five hours under vacuum. To determine the final thickness, we
measure the film's weight and size. We validated this method by comparison with an optical
technique using interference fringes.

\begin{figure}[b]
    \centering
    \includegraphics[width=7cm]{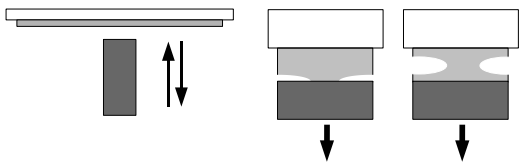}
    \caption{\textit{Left side:} Schematic view of the ''probe tack``
    experiment. \textit{Right side:} Interfacial crack propagation and bulk
    deformation mechanisms.}
    \label{fig:scheme_mutack}
\end{figure}

We perform tensile deformation tests using a home built ''probe tack`` set up with good resolution
and visualization capabilities \cite{Josse_JADHES_2004}. It mainly consists of a flat circular
steel probe that is brought into contact and debonded from a soft viscoelastic film with controlled
speed, see figure \ref{fig:scheme_mutack}. During the test, the probe displacement and the normal
force on the probe are measured. We also visualize the debonding process from above with a camera
mounted on a microscope to gain qualitative insight into the debonding mechanisms. The probe has a
radius $R = 3mm$ and is made of polished stainless steel.\\

\begin{figure}
    \includegraphics[width=8.6cm]{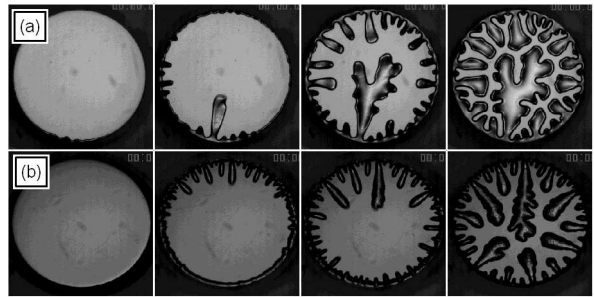}
    \caption{Formation of air fingers in the elastic and viscoelastic case: interfacial crack
    propagation (a) and bulk deformation (b).}
    \label{fig:pictures_finger}
\end{figure}

\begin{figure}
    \centering
    \includegraphics[width=7cm]{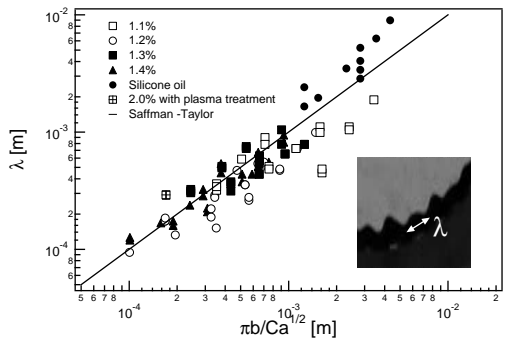}
    \caption{In the viscoelastic case, the wavelength scales linearly
    with the thickness and is inversely proportional to the square root
    of the capillary number.}
    \label{fig:lambda_vs_controlparavisc}
\end{figure}

\textsc{Experimental} -- The parameters varied in our experiments, besides the viscoelastic
properties, are the layer thickness $b$ and the debonding speed $v$. Typical values are $b = 50 -
500\mu m$ and $v=1-200\mu m/s$. During a typical experiment, air penetrates from the edge of the
confined layer. It can penetrate either in the bulk, followed by a strong deformation and the
subsequent formation of thin "bridges" (fibrils) between the probe and the glass slide, or at the
interface between the probe surface and the polymer film, leading to a fast debonding by
interfacial crack propagation. In both cases, we observe the destabilization of the initially
circular debonding line by undulations and the subsequent propagation of air fingers. We
characterize the emerging patterns by determining the finger number $n$ at the moment the first
undulations are observable, see inset of figure \ref{fig:lambda_vs_controlparavisc}, and calculate
a wavelength $\lambda = 2\pi R/n .$ Initially a destabilizing wavelength can be clearly defined,
but as the time and debonding process go on, highly non-linear patterns are evolving, showing
features like \textit{side branching} and \textit{tip splitting}, see figure
\ref{fig:pictures_finger}. In the present study we restrict our interest to the analysis of the
linear destabilization process at the onset.\\

\textsc{Results and Discussion} -- We characterize here in more detail the two cases of interfacial
and bulk mechanisms introduced above. Although the patterns look quite similar in the top view
pictures on figure \ref{fig:pictures_finger} (a) and (b), two different mechanisms are at their
origin.

\begin{figure}
    \includegraphics[width=8.6cm]{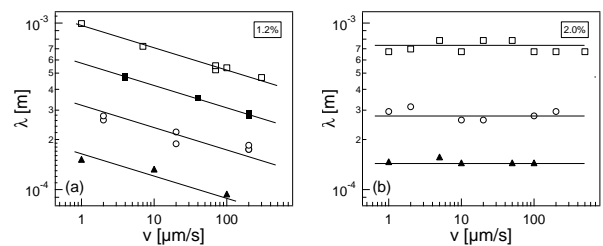}
    \caption{\textit{Left side, viscoelastic case:} $\lambda$ depends on $b$ and $v$.
    $\square = 350\mu m, \blacksquare = 230\mu m, \circ= 120\mu m, \blacktriangle = 60\mu m.$
    \textit{Right side, elastic case:} $\lambda$ depends only on $b$.
    $\square = 300\mu m, \circ = 130\mu m, \blacktriangle = 70\mu m.$ Black lines are a guide for the eye.}
    \label{fig:lambda_vs_v}
\end{figure}

In the case of the \textbf{viscoelastic regime} characterized by fibrillation and a bulk
deformation mechanism, the pattern formation is sensitive to both the initial film thickness and
the debonding speed for a given material. As the wavelength decreases with the debonding speed and
increases linearly with the initial film thickness (figure \ref{fig:lambda_vs_v}(a)), one can
attempt to compare $\lambda$ to the classical \textit{Saffman - Taylor} (\textit{ST}) or viscous
fingering instability \cite{Saffman_RROCRSOC_1958,Paterson_JFLUIDMECH_1981} predicting by linear
stability analysis
\begin{equation}
    \label{eq:lambda_ST}
    \lambda = \pi b / \sqrt{C\!a}\ .
\end{equation}

$C\!a =U\eta / \sigma$ is the dimensionless capillary number comparing viscous to capillary forces,
$\eta$ the viscosity, $\sigma = 20mN/m$ the surface tension between PDMS and air, and $U$ denotes
the radial velocity of the circular interface. Presuming an incompressible fluid and therefore
volume conservation, $U = Rv/2b$ for a Newtonian fluid. To adapt this prediction to the case of
viscoelastic materials, we replace the Newtonian viscosity with a complex viscosity $|\eta^\star|$
defined as $G^\star / \omega$. $|\eta^\star|$ depends on the frequency, estimated for each of our
experiments following $\omega = 2\pi U/b$.

Figure \ref{fig:lambda_vs_controlparavisc} shows a good quantitative agreement between the
\textit{ST} prediction and our data, despite some scattering. The limit of a purely viscous liquid
is represented by the dark full spots obtained for Newtonian silicone oils \cite{Data_oil}.
Surprisingly, the \textit{ST} prediction holds going from the viscous limit up to highly
non-Newtonian viscoelastic materials above the gel point.

\begin{figure}
    \centering
    \includegraphics[width=7cm]{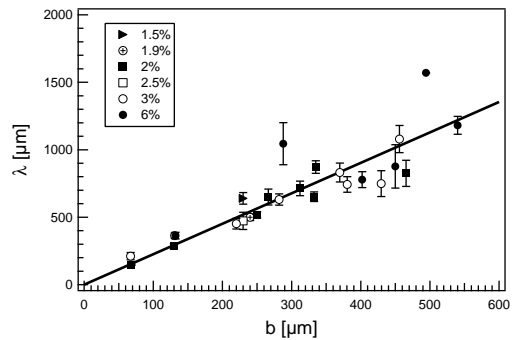}
    \caption{In the case of interfacial crack propagation, the
    wavelength only scales with the thickness $b$.
    The solid line is a straight line fit to the data yielding
    $\lambda = 2.3b$.}
    \label{fig:lambda_vs_thickness}
\end{figure}

 The second case we investigated, the \textbf{elastic regime}, is characterized by interfacial crack
propagation. The linear wavelength does not depend on the debonding speed over three decades, see
figure \ref{fig:lambda_vs_v}(b). The dependence on the debonding speed is a quantitative criterion
to decide which regime an experiment belongs to. Figure \ref{fig:lambda_vs_thickness} shows that
$\lambda$ depends only on the initial film thickness $b$ over three orders of magnitude of the
elastic modulus ($1kPa\lesssim G' \lesssim 0.5MPa$). These results are in qualitative agreement
with theoretical predictions and with experimental observations in a slightly different geometry
\cite{Ghatak_all}. Linear stability analysis has also been done by Adda-Bedia and Mahadevan
\cite{Adda-Bedia_PROCRSOCA_2006}. Considering the case of static peeling, they take into account
the bending stiffness of the flexible cover plates used for the peeling tests and the finite film
thickness. They calculate a critical confinement above which shear deformations are more beneficial
for the system's energy than normal deformations, leading to undulations. The confinement parameter
$\alpha$ is defined as $(D / E b^3)^{1/3}$, $D$ being the bending stiffness of the cover plate, $E$
the film's elastic modulus and $b$ the film thickness. The critical value $\alpha_c \simeq 21$ is
in good agreement with experiments by Ghatak \textit{et al} who find $\alpha_c \simeq 18$. We
compare our experiments to these results by considering the bending stiffness of our microscopic
glass slides. With $D \simeq 70 Nm$ for a glass slide of $b = 2mm$, we find $\alpha > 70$ for all
our experiments, thus we place ourselves always in the regime of an unstable crack front. The
critical wavelength calculated in \cite{Adda-Bedia_PROCRSOCA_2006} $\lambda_c \simeq 3.4 b$ scales
only with the film thickness and is independent of all material parameters. Our result $\lambda =
2.3b$ is in good quantitative agreement with theory. Deviations might be due to the fact that
calculations are done for $\alpha = \alpha_c$ whereas our experiments are placed far beyond the
critical value.

\begin{figure}
    \centering
    \includegraphics[width=7cm]{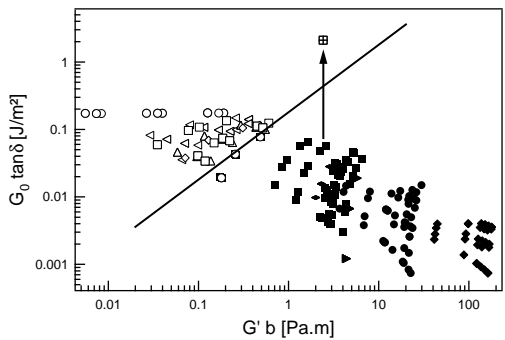}
    \caption{Open symbols represent bulk deformation and black full symbols interfacial crack
    propagation. Experiments right at the transition can show both mechanisms due to
    fluctuations in the sample preparation.}
    \label{fig:parameterspace}
\end{figure}

A surprising result of our work is the very abrupt change in the debonding behavior: our
experiments always fall into the elastic or viscoelastic regime without experiencing a transition
regime. The appropriate parameter to describe the transition between interfacial and bulk
mechanisms in the case of an elastic rubber has been proposed to be $\mathcal{G}_c / E b$
\cite{Webber_PHYSREVE_2003}. The critical energy release rate $\mathcal{G}_c$ is a measure for the
energy one has to provide to the system to make an interfacial crack move. $E b$ represents the
elastic energy necessary to deform the bulk of a sample of thickness $b$ with elastic modulus $E$.
For a viscoelastic material, $\mathcal{G}_c$ can be divided into a constant component
$\mathcal{G}_0$, the threshold fracture energy, and a dissipation term depending on crack velocity.
It has been proposed \cite{Maugis_JPHYSD_1978} that the dissipation term should be proportional to
$\tan \delta = G''/G'$. Hence approximating $\mathcal{G}_c\sim\mathcal{G}_0 \tan \delta$ and
substituting into $\mathcal{G}_c / E b$ yields for soft viscoelastic layers a new parameter
$(\mathcal{G}_0 \tan \delta) / (G' b)$ depending only on the linear rheological properties  and
$\mathcal{G}_0$ \cite{Carelli_JofAdh_2007}. For cases where the energy cost to propagate a crack is
high, bulk mechanisms are expected, while interfacial crack propagation should be observed when the
elastic deformation of the layer requires high energy. This is well presented by plotting the
parameter space spanned by $(\mathcal{G}_0 \tan \delta)$ and $(G' b)$, see figure
\ref{fig:parameterspace}. Full symbols indicate interfacial, open symbols bulk mechanisms.

Following the theory it should be possible to switch between interfacial and bulk mechanism by
changing $\mathcal{G}_0$. We performed an exemplary experiment replacing the steel probe
($\mathcal{G}_0\!\simeq\!0.1J/m^2$) by a glass surface previously subjected to plasma treatment,
increasing $\mathcal{G}_0\! \simeq\! 15J/m^2$ considerably. We estimated $\mathcal{G}_0$ by
measuring the work of adhesion for the fully cured PDMS performing a tack test at low debonding
speed. We were indeed able to change the debonding mechanism from interfacial to bulk behavior for
a sample with $2\%$ of cross linker. This experiment is represented by the symbol $\boxplus$ on
figure \ref{fig:parameterspace}. Furthermore, changing $\mathcal{G}_0$ changes the wavelength which
is now well described by the \textit{ST}
prediction,see $\boxplus$ on figure \ref{fig:lambda_vs_controlparavisc}.\\

\textsc{Conclusion} -- We present in this Letter for the first time a systematic study of the
transition between bulk deformation mechanisms and interfacial crack propagation during tensile
tests on thin layers of viscoelastic materials with properties going from a viscous liquid to an
elastic solid. In both cases, we characterize the emerging fingering patterns quantitatively
following theoretical predictions. The transition we observe is very sharp without experiencing an
intermediate regime. We propose a possible empiric parameter that allows to draw a mechanism map
spanned by the parameters $\mathcal{G}_0 \tan \delta$ and $G' b$ separating nicely the different
mechanisms and allowing therefore to predict the debonding behavior of our system.

We thank Mokhtar Adda-Bedia, Arezki Boudaoud and Dominic Vella (LPS-ENS) for stimulating
discussions and Guylaine Ducouret (PPMD-ESPCI) and Alexis Prevost (LPS-ENS) for help with the
sample characterization.


\end{document}